\documentclass[12pt]{article}

\usepackage{graphicx}
\usepackage{cite}

\begin{document}

\hoffset = -1truecm \voffset = -2truecm \baselineskip = 8 mm

\title{\bf  Nucleon spin structure III: Origins of the generalized Gerasimov-Drell-Hearn sum rule}

\author{
 {\bf Wei Zhu} and {\bf Jianhong Ruan}\\
\\
\normalsize Department of Physics, East China Normal University,
Shanghai 200062, P.R. China \\}

\date{}

\newpage

\maketitle

\vskip 3truecm

\begin{abstract}

    The generalized Gerasimov-Drell-Hearn (GDH) sum rule is understood based
on the polarized parton distributions of the proton with the higher
twist contributions. A simple parameterized formula is proposed to
clearly present the contributions of different components in proton
to $\Gamma_1^p(Q^2)$. We find that the contribution of quark
helicity to $\Gamma_1^p(Q^2)$ is almost constant ($\sim 0.123$);
the twist-4 effect dominates the suppression of $\Gamma_1^p(Q^2)$ at
$Q^2<3GeV^2$, while the twist-6 effect arises a dramatic change of
$\Gamma_1^p(Q^2)$ at $Q^2<0.3 GeV^2$, it implies a possible extended
objects with size $0.2-0.3~ fm$ inside the proton.

\end{abstract}

PACS number(s): 12.38.Cy, 12.38.Qk, 12.38.Lg

$keywords$:  Nucleon spin structure

\newpage
\begin{center}
\section{Introduction}
\end{center}

    The structure of nucleon is an important open question.
The perturbative QCD (pQCD) at high energy has been proved a
successful approach to describe the phenomena in terms of elementary
quark and gluon constituents. On the other hand, hadronic processes
at low transfer momentum are usually described by using meson and
baryon degrees of freedom. Unfortunately, the transition from
quark-gluon to hadron degrees of freedom  remains shrouded in
mystery.

    There is particular interest in the first moment $\Gamma_1(Q^2)=\int_0^1 dxg_1(x,Q^2)$
of the spin structure functions $g_1(x,Q^2)$, which has been
measured from high $Q^2$ down to $\sim 0~ GeV^2$. The goal of
obtaining universal expressions describing $\Gamma_1(Q^2)$ at any
$Q^2$ is an attractive task for both theoretical and
phenomenological point of view. In theory, $\Gamma_1(0)$ is
constrained by the Gerasimov-Drell-Hearn (GDH) sum rule [1,2]

$$I_1(0)=\lim_{Q^2\rightarrow 0}\frac{2M^2}{Q^2}\Gamma_1(Q^2)=
-\frac{\kappa^2}{4}\sim -0.8,\eqno(1.1)$$where $\kappa$ is the
anomalous magnetic moment of the nucleon. On the other hand, the
Bjorken sum rule [3] says

$$\lim_{Q^2\rightarrow \infty}[\Gamma_1^p(Q^2)-\Gamma_1^n(Q^2)]=\frac{1}{6}\left\vert\frac{g_A}{g_V}\right\vert,\eqno(1.2)$$
this ratio is known very accurately [4]: $g_A/g_V=-1.2695\pm0.0029$.

The connection of the two sum rules by means of the generalized GDH
sum rule is(For an overview see, for example Ref. [5])

$$I_1(Q^2)=\frac{2M^2}{Q^2}\Gamma_1(Q^2),\eqno(1.3)$$which
allows us to study the transition between the perturbative partonic
structure and non-perturbative hadronic picture of nucleon in
lepton-nucleon scattering processes. The data show that this sum
rule at low $Q^2<1GeV^2$ changes dramatically and exceeds the
variation bound at higher $Q^2$, which has been parameterized (but
not explanation) in [6]. The explanation of the generalized GDH sum
rule is an active subject.  For example, the phenomenological
constituent quark model [7], the vector meson dominate (VMD) model
[8], the resonance contributions [9], the chiral perturbation theory
($\chi$PT) [10] are used to understand the generalized GDH sum rule.

    The purpose of this work is try to expose the
partonic structure in the generalized GDH sum
rule. According to the parton model, the polarized structure
functions of  nucleon at $Q^2>$ a few $GeV^2$ and the leading
order (LO) approximation are defined by the parton distributions as

$$g_1^{DGLAP}(x,Q^2)=\sum_ie^2_i[\delta q_i(x,Q^2)+\delta \overline{q}_i(x,Q^2)],\eqno(1.4)$$
where the polarized parton distributions obeys the DGLAP evolution
equation [11].  In our previous works [12,13], the DGLAP equation
with the parton recombination corrections [14,15] are used to derive
the unpolarized and polarized parton distributions starting from
simple initial valence quark distributions at $Q^2\sim 0.064GeV^2$.
Our numerical calculations show that the predicted parton
distributions consist with the data well both for polarized and
unpolarized data at $Q^2>1~GeV^2$. Considering the nonperturbative
corrections to the spin structure function at low $Q^2$ becoming
important, Eq. (1.4) in the full $Q^2$ range should be generalized
to

$$g_1^p(x,Q^2)=g_1^{DGLAP+ZRS}(x,Q^2)+g_1^{VMD}(x,Q^2)+g_1^{HT}(x,Q^2),\eqno(1.5) $$
where $g_1^{DGLAP+ZRS}(x,Q^2)$ is the solutions of the DGLAP
equation with the parton recombination (ZRS) corrections at the
$LL(Q^2)$ approximation [15], while $g_1^{VMD}(x,Q^2)$ is the
contributions of the vector meson in virtual photon via the vector
meson dominated (VMD) model [16], and $g_1^{HT}(x,Q^2)$ is the
higher twist corrections. The corresponding first moment of $g_1^p$ is

$$\Gamma_1^p(Q^2)=\Gamma_1^{DGLAP+ZRS}(Q^2)+\Gamma_1^{VMD}(Q^2)+\Gamma_1^{HT}(Q^2).\eqno(1.6)$$

    Since we have known the contributions of
$g_1^{DGLAP+ZRS}$ and $g_1^{VMD}$, one can expose the properties of
$\Gamma_1^{HT}(Q^2)$ after subtracting these two contributions from
the experimental data about $\Gamma_1^p(Q^2)$. This opens a window
to visit higher twist effects at low $Q^2$ in the nucleon structure.
We find that (i) the contributions of quark helicity to
$\Gamma_1^p(Q^2)$ are almost constant ($\sim 0.123$); (ii) the
negative twist-4 effect dominates the suppression of
$\Gamma_1^p(Q^2)$ at $Q^2<3GeV^2$, while the twist-6 effect becomes
positive at $Q^2\sim 1~GeV^2$ and it arises a dramatic change of
$\Gamma_1^p(Q^2)$ at $Q^2<0.3 GeV^2$, these effects can be described
by a simple parameterized form; (iii) the slope of the higher twist
coefficient suddenly  enhance at $Q^2\sim 1~GeV^2$, which is near an
estimation of the constituent quark scale $0.2\sim 0.3~fm$.

    We will give a review
of the structure functions in full $Q^2$ range based on the partonic
picture in Sec. 2, where the contributions of parton helicity and
the VMD mechanism in our previous works are summarized. Then we
extract the twist-4 and 6 corrections to $\Gamma_1^p(Q^2)$ in Sec.
3, which expose the interesting change of the partonic structure of
proton  at the low $Q$-scales. The discussions and a summary are
presented in Sec. 4.

\newpage
\begin{center}
\section{Spin structure functions in the full $Q^2$ range}
\end{center}

    We have detailed the first two terms of Eq. (1.5) in our previous two works [13,16].
$g_1^{DGLAP+ZRS}$ are the contributions of parton helicities obeying
the DGLAP equation with the parton recombination corrections
[14,15], where the input valence quark distributions at
$\mu^2=0.064GeV^2$ have been fixed through a global fitting:

$$xu_v(x,\mu^2)=24.30x^{1.98}(1-x)^{2.06},\eqno(2.1)$$

$$xd_v(x,\mu^2)=9.10x^{1.31}(1-x)^{3.80}, \eqno(2.2)$$
for unpolarized parton distributions, and

$$\delta u_v(x,\mu^2)=40.3x^{2.85}(1-x)^{2.15},\eqno(2.3)$$

$$\delta d_v(x,\mu^2)=-18.22x^{1.41}(1-x)^{4.0},\eqno(2.4)$$
for polarized parton distribution. The input distributions
Eqs.(2.1-2.4) neglect the contributions of asymmetry sea quark
contributions in this work. We consider these corrections to
$\Gamma_1$ are small.

    Using Eqs. (2.1-2.4) we obtain $\Gamma^{DGLAP+ZRS}(Q^2)\simeq
0.123$, which seems insensitive to the complicated
behavior of $g_1^p$ at small $x$ [16].

    Note that $1/\mu= 0.78 fm$ consists with a typical proton scale
$0.8\sim 1~fm$. We assume that $\mu$ is a minimum average transverse
momentum of the partons in the proton due to the uncertainty
principle. Thus, we assume that the $Q^2$-dependence of parton
distributions are freezed at $Q^2<\mu^2$. Using this assumption we
avoid the un-physical singularities at $Q\sim \Lambda_{QCD}$. The
dashed curve in Fig. 1 is our predicted $\Gamma_1^{DGLAP+ZRS}(Q^2)$.
We find that it is almost irrelevant to $Q^2$.

    The second term of Eq. (1.5) is the the contributions from the
interaction of $\rho$- and $\omega$-mesons in the virtual photon
with the proton and they are described by the vector meson dominance
(VMD) model [17],

$$xg_1^{VMD}(x,Q^2)=\frac{Q^2}{8\pi}\sum_v\frac{m^4_v\Delta\sigma_{vp}(s)}{\gamma_v^2(Q^2+m^2_v)^2}. \eqno(2.5)$$
In this formula the constants $\gamma_v$ are determined by the
leptonic widths of the vector mesons and $m_v$ denotes the mass of
$\rho$ and $\omega$ at $Q^2<1 GeV^2$. The cross-sections $\Delta
\sigma_{v}(s)$ are the unknown cross sections for the scattering of
polarized mesons and nucleons. s is the CMS energy squared for the
$\gamma p$ collision and $s\sim Q^2/x$. In work [16] we assume

$$g_1^{VMD}(x,Q^2)=B\frac{(m^2_V)^{2-\lambda+\epsilon}Q^2)^{\lambda}}{(Q^2+m^2_V)^2}[(\frac{x}{x_0})^{-\lambda+\epsilon}\theta
(x_0-x)+\frac{\ln^4x}{\ln^4x_0}\theta(x-x_0)],\eqno(2.6)$$
$\lambda=1$ and $\epsilon$ is a small positive parameter due to the
requirement of integrability of $g_1^p$ at $x\rightarrow 0$.
 Thus, we have

$$\Gamma_1^{VMD}(Q^2)=\int^1_0dxg_1^{VMD}(x,Q^2)\simeq A\frac{m^2_vQ^2}{(Q^2+m^2_v)^2},
\eqno(2.7)$$where we take $A=0.162$, which corresponds to
$\epsilon=0.018$. The solid curve in Fig. 1 is
$\Gamma_1^{DGLAP+ZRS}(Q^2)+\Gamma_1^{VMD}(Q^2)$, it shows a weaker
$Q^2$-dependence at $Q^2>1GeV^2$.

    Comparing the solid curve with $\Gamma_1^p(Q^2)$ data [20-22] in Fig. 1,
one can expect that the higher twist corrections $g_1^{HT}$ play a
significant role at low $Q^2$ to the general GDH sum rule. We will
detail them in next section.

\newpage
\begin{center}
\section{Higher twist contributions to the GDH sum rule}
\end{center}

    According to the operator product expansion (OPE), the
appearance of scaling violations at low $Q^2$ is related to the
higher twist corrections to moments of structure functions.
Higher twists are expressed as matrix elements of operators
involving nonperturbative interactions between quarks and gluons.
The study of higher twist corrections gives us a direct insight into
the nature of long-range quark-gluon correlations. The higher twist
corrections to $g_1$ have several representations. The initial
parton recombination corrections have been examined in our works
[12,13]. The VMD model was used to simulate a special higher twist effect
in our previous work [16]. In this work, we will try to expose the
remaining power suppression corrections to $\Gamma_1^p$. For this
sake, we make $\Gamma_1^p(Q^2)$ (i.e., the data points [28-20] in
Fig. 1)-[$\Gamma_1^{DGLAP+ZRS}(Q^2) +\Gamma_1^{VMD}(Q^2)$] (i.e.,
the solid curve in Fig. 1). Figure 2 shows such a result at
$Q^2>0.2~GeV^2$, which has been smoothed with minimum
$\chi^2/D.o.f.$.

    To expose the possible physical information of the curve in Fig. 2, according to QCD
operator product $1/Q^2$-expansion,

$$\Gamma_1^{HT}(Q^2)=\sum_{i=2}^{\infty}\frac{\mu_{2i}(Q^2)}{Q^{2i-2}},\eqno(3.1)$$ we take first three approximations

$$\Gamma_1^{HT(4)}(Q^2)=\frac{\mu_4(Q^2)}{Q^2},\eqno(3.2)$$

$$\Gamma_1^{HT(4+6)}(Q^2)=\frac{\mu_6(Q^2)+\mu_4(Q^2)Q^2}{Q^4},\eqno(3.3)$$

$$\Gamma_1^{HT(4+6+8)}(Q^2)=\frac{\mu_8(Q^2)+\mu_6(Q^2)Q^2+\mu_4(Q^2)Q^4}{Q^6}.\eqno(3.4)$$

    Then we plot the curves $Q^2\Gamma_1^{HT(4)}(Q^2)$, $Q^4\Gamma_1^{HT(4+6)}(Q^2)$ and  $Q^6\Gamma_1^{HT(4+6+8)}(Q^2)$
in Fig. 3. There are following interesting properties of these
results:

 (i) $Q^6\Gamma_1^{HT(4+6+8)}(Q^2)\rightarrow 0$, if $Q^2\rightarrow 0$. This implies that $\mu_8$ vanishes if it
is independent of $Q^2$. Therefore, $\Gamma_1^{HT(4+6)}(Q^2)$ is an
appropriate approximation.

 (ii) Three curves in Fig. 3 cross at a same point $Q^2\sim 1 GeV^2$.
Particularly, the intercept $\mu_6$ of the line, which cuts
$Q^4\Gamma_1^{HT(4+6)}(Q^2)$ suddenly changes its sing from -0.03 at
$Q^2>1~GeV^2$ to 0.01 at $Q^2<1~GeV^2$. This result exposes that the
correlation among partons in the proton has an obvious change near
$Q\sim 1 GeV$.

 (iii) Although $\Gamma^{HT}_1(Q^2)\sim\Gamma_1^{HT(4)}(Q^2)=-0.05/Q^2$ is an acceptable
approximation at $Q^2>1.5~GeV^2$, the correct choose is
$\Gamma^{HT}_1(Q^2)=\Gamma_1^{HT(4+6)}(Q^2)$. We use

$$\Gamma_1^{HT(4+6)}(Q^2)=\frac{\mu_4}{Q^2+\epsilon^2}+\frac{\mu_6}{Q^4+\epsilon^4}~~at~Q^2<0.3 GeV^2\eqno(3.5)$$ to fit the data
at $Q^2<0.3 GeV^2$, where we add a parameter $\epsilon$ to remove
the unnatural singularity at $Q^2=0$. The value of $\epsilon$ is
sensitive to $I(0)$. We find that $\mu_4=-0.154$, $\mu_6=0.037 $ and
$\epsilon^2=0.32 GeV^2$. On the other hand, the negative twist-4
effect dominates the suppression of $\Gamma_1^p(Q^2)$ at
$Q^2<3GeV^2$, while the twist-6 effect change its sing at $Q^2=1
GeV^2$ and its positive effect arises a dramatic change of
$\Gamma_1^p(Q^2)$ at $Q^2<0.3 GeV^2$.

        In summary,

$$\Gamma_1^p(Q^2)=0.123+0.162\frac{m^2_vQ^2}{(Q^2+m^2_v)^2}+\Gamma_1^{HT}(Q^2)\eqno(3.6)$$where
the HT contributions are

$$\Gamma_1^{HT}(Q^2)=\left\{
\begin{array}{ll}
-\frac{0.043/M^2}{Q^2}-\frac{0.03/M^2}{Q^4}~~at~Q^2>1 GeV^2\\
-\frac{0.0844/M^2}{Q^2}+\frac{0.0114/M^2}{Q^4}~~at~0.3<Q^2<1 GeV^2\\
-\frac{0.156/M^2}{Q^2+0.32}+\frac{0.037/M^2}{(Q^2+0.32)^2}~~at~Q^2<0.3~GeV^2
\end{array}
\right.\eqno(3.7)$$where $M^2=1GeV^2$ is the Bore1 parameter used in
the sum rule. We present the comparison of our
$\Gamma_1^p(Q^2)$ with the data [18-20] in Fig. 4. The corresponding
$I_1^p(Q^2)$ using Eqs. (1.3), (3.6) and (3.7) is presented in Fig.
5.

\newpage
\begin{center}
\section{Discussions and summary}
\end{center}

    (i)  The parton-hadron duality was first noted by Bloom and
Gilman [21] in deep inelastic scattering (DIS) and has been
confirmed by many measurements. At low energies (or intermediate
Bjorken variable $x$ and low $Q^2$) DIS reactions are characterized
by excitation of nucleon resonances; while at high virtuality such
processes have a partonic description. The smooth high-energy
scaling curve essentially reproduces the average of the resonance
peaks seen at low energies. Burkert and Ioffe [6] indicated that the
contribution of the isobar $\Delta(1232)$ electroproduction at small
$Q^2$ can describe the general GDH sum rule, and they gave

$$\frac{\mu_4}{M^2}=-0.056\sim -0.063, at ~Q^2=0.3\sim 0.8 ~GeV^2\eqno(4.1)$$

$$\frac{\mu_6}{M^2}=0.010\sim 0.011, at~ Q^2=0.3\sim0.8 ~GeV^2, \eqno(4.2)$$ which
are compatible with our prediction Eq. (3.7).

    Our results in Fig.3  indicates the slope of the twist
coefficient $\mu_4(Q^2)$ in Eq. (3.7) suddenly  enhance at $Q^2\sim
1~GeV^2$. It implies that the correlation among partons become
stronger at scale $\sim 0.2~fm$. We noted that Petronzio1, Simula and
Ricco [22] reported that the inelastic proton data obtained at
Jefferson Lab exhibit a possible extended objects with size of
$\simeq 0.2-0.3~ fm$ inside the proton.

    As we have shown that the twist $N>6$ corrections are neglected at low
$Q^2$. This supports our generalized leading order (GLO)
approximation [13,16]: if the lower order contributions are
compatible with the experimental data at low $Q^2$, one can image
that the higher order corrections are cancelable each other.

    In summary, we assume that the parton
distributions are still available at $Q^2< 1~GeV^2$. Using the DGLAP
equation with the parton recombination corrections, we indicate that
the contributions of $Q^2$-dependent parton distributions to the
lowest moment of the spin-dependent proton structure function are
almost constant in full $Q^2$ range. In the meantime, the
contributions of a VMD-type nonperturbative part are isolated. After
removing the above two contributions from the existing experimental
data for $\Gamma_1^p(Q^2)$, the higher twist power corrections present
their interesting characters: parton correlations at $Q^2\sim
1~GeV^2$ and $\sim 0.3~GeV^2$ show two bend points, where the
twist-4 effect dominates the suppression of $\Gamma_1^p(Q^2)$ at
$Q^2<3GeV^2$, while the twist-6 effect arises a dramatic change of
$\Gamma_1^p(Q^2)$ at $Q^2<0.3 GeV^2$. Within the analytic of these
results, we are able to achieve a rather good description of the
data at all $Q^2$ values using a simple parameterized form of
$\Gamma_1^p(Q^2)$.

\begin{figure}[htp]

\vskip 3cm
\centering
\includegraphics[width=0.7\textwidth]{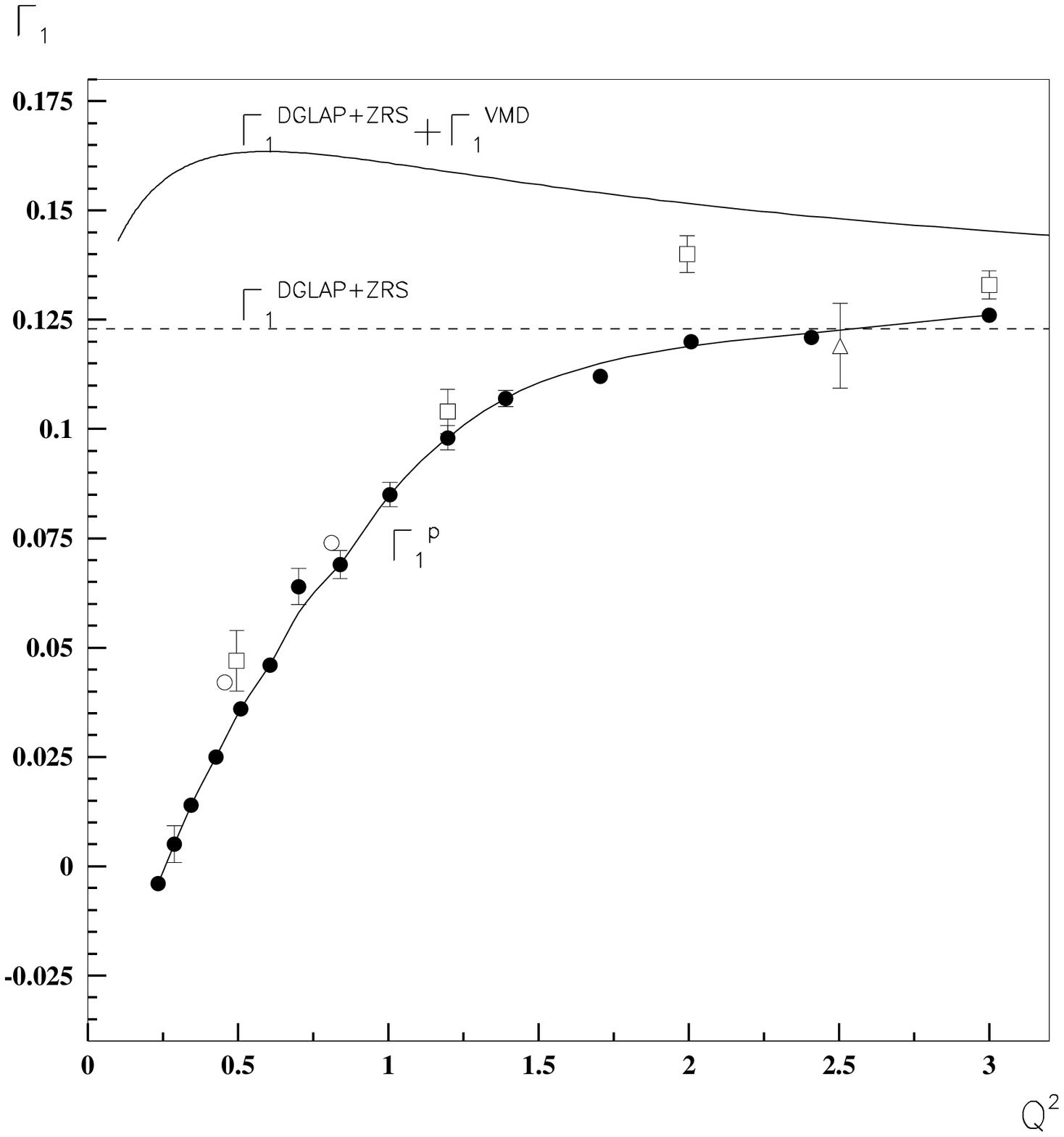}
\vskip 1cm \caption{ Contribution of quark helicity
$\Gamma_1^{DGLAP+ZRS}(Q^2)$ (dashed curve) and combining VMD
contribution $\Gamma_1^{DGLAP+ZRS}(Q^2)+\Gamma_1^{VMD}(Q^2)$ (solid
curve). Data are from from Hermes experiment at DESY [18], the E143
experiment at SLAC [19] and the EG1a experiment using the CLAS
detector at JLab [20].} \label{fig1}
\end{figure}

\newpage

\begin{figure}[htp]

\vskip 3cm
\centering
\includegraphics[width=0.7\textwidth]{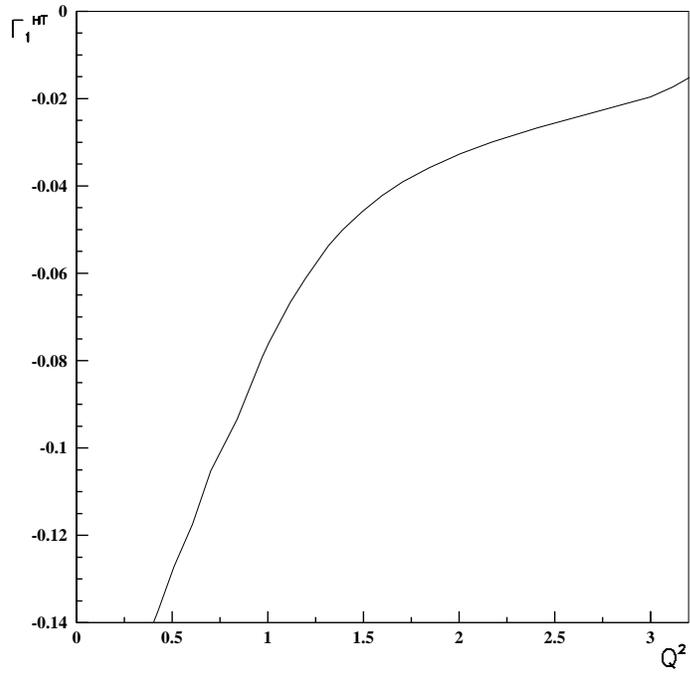}
\vskip 1cm
\caption{ Contribution of higher twist $\Gamma_1^{HT}(Q^2)$ (smoothed
curve) is taken from
data-[$\Gamma_1^{DGLAP+ZRS}(Q^2)+\Gamma_1^{VMD}(Q^2)]$.} \label{fig2}
\end{figure}

\newpage

\begin{figure}[htp]

\vskip 3cm
\centering
\includegraphics[width=0.7\textwidth]{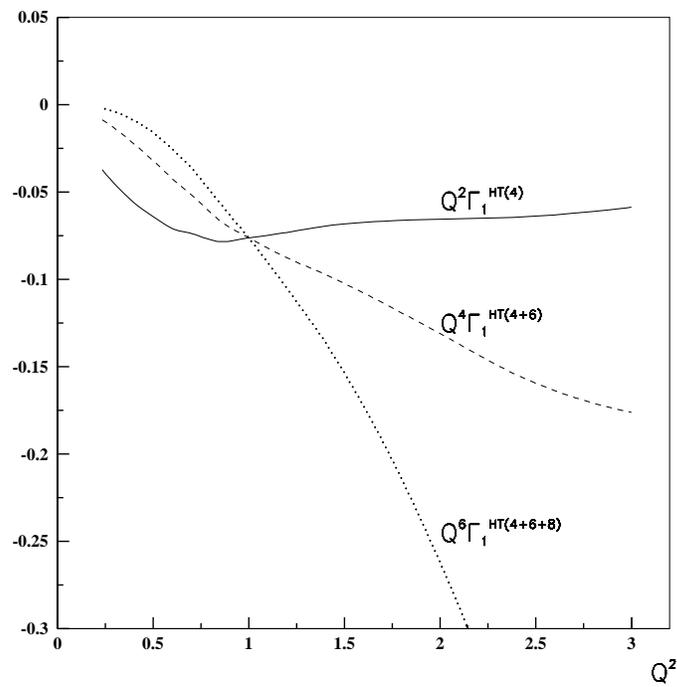}
\vskip 1cm
\caption{ Three different analysis of the higher twist contributions.} \label{fig3}
\end{figure}

\newpage

\begin{figure}[htp]

\vskip 3cm
\centering
\includegraphics[width=0.7\textwidth]{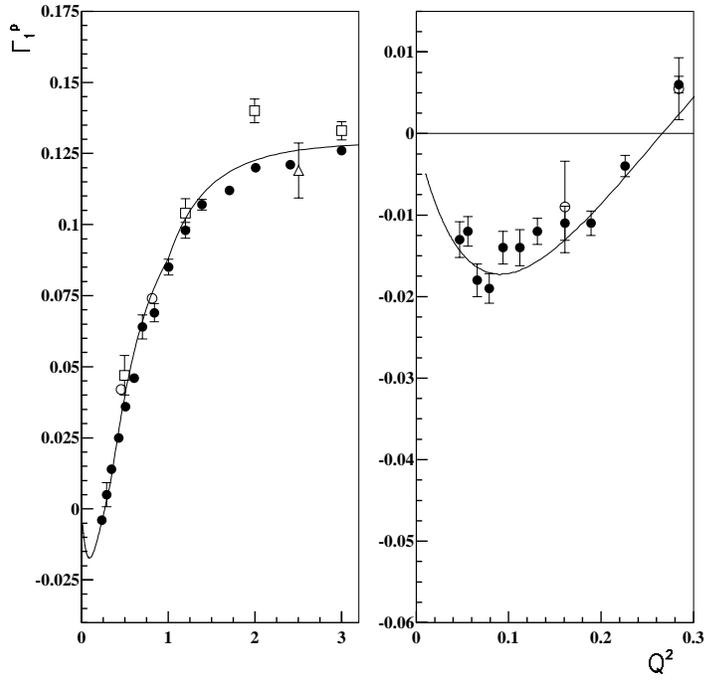}
\vskip 1cm
\caption{The $Q^2$ dependence of $\Gamma_1^p(Q^2)$ calculated by Eqs.
(3.6) and (3.7). The data are taken from [18-20].} \label{fig4}
\end{figure}

\newpage

\begin{figure}[htp]

\vskip 3cm
\centering
\includegraphics[width=0.7\textwidth]{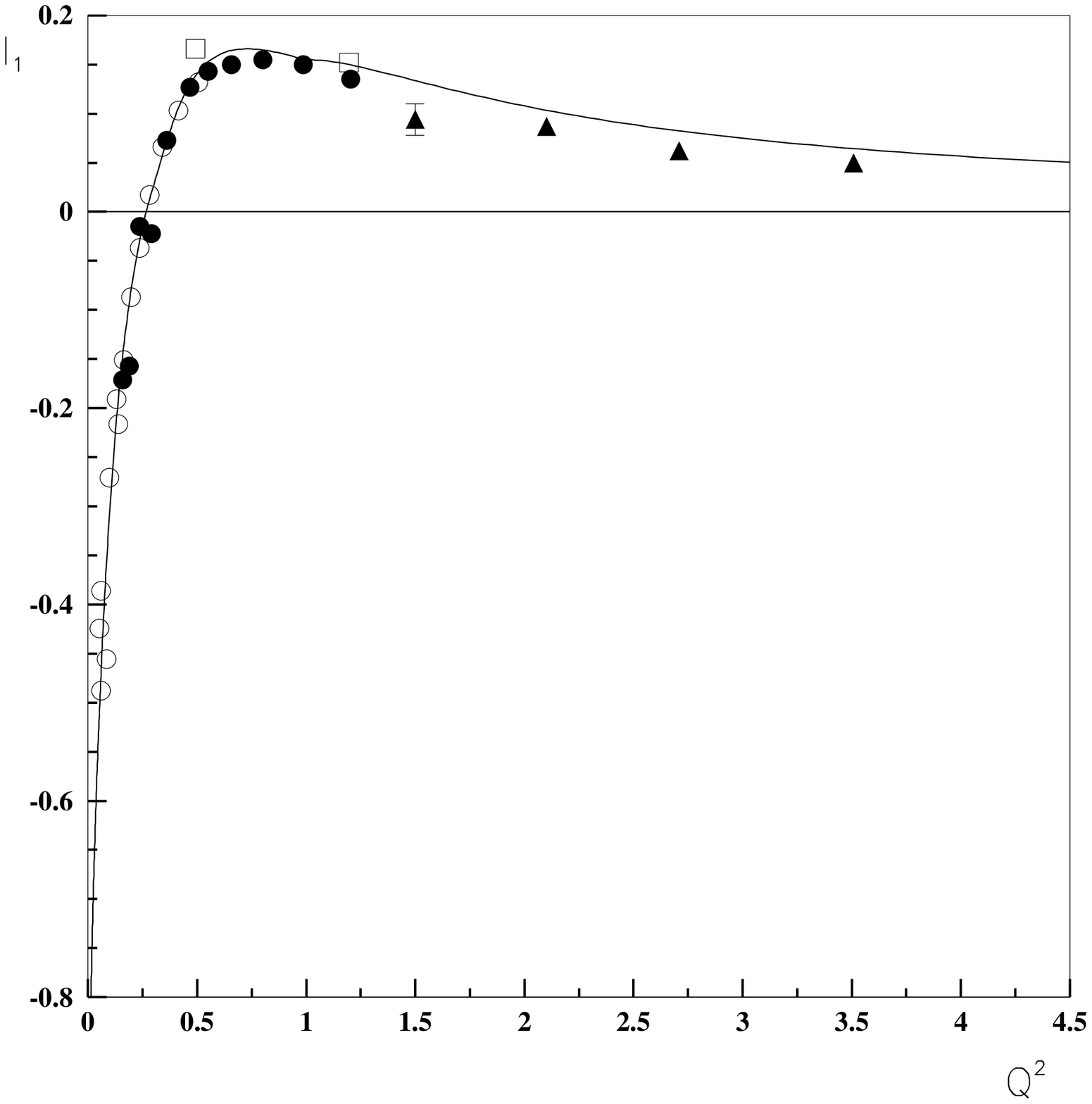}
\vskip 1cm
\caption{ The $Q^2$ dependence of $I_1^p(Q^2)$ calculated by Eqs.
(1.3), (3.6) and (3.7). The data are taken from [18-20].} \label{fig5}
\end{figure}


\newpage
\begin{thebibliography}{99}

\bibitem{1} S.B. Gerasimov, Sov. J. Nucl. Phys. $\bf{2}$, 430 (1966).

\bibitem{2} S.D. Drell and A.C. Hearn, Phys. Rev. Lett. $\bf{16}$, 908
(1966).

\bibitem{3} J.D. Bjorken, Phys. Rev. $\bf{148}$, 1467 (1966).

\bibitem{4} Particle Data Group, S. Eidelman et al., Phys. Lett. $\bf{B592}$, 1 (2004).

\bibitem{5} D. Drechsel, S.S. Kamalov and L. Tiator, Phys. Rev. $\bf{D63}$, 114010 (2001)
hep-ph/0008306.

\bibitem{6} M. Anselmino, B.L. Ioffe and E. Leader, 1989, Sov. J. Nucl.
Phys. $\bf{49}$, 136 (1989); V.D. Burkert and B.L. Ioffe,  Phys.
Lett. $\bf{B296}$, 223 (1992);  J. Soffer and O.V. Teryaev, Phys.
Rev. Lett. $\bf{70}$, 3373 (1993); J. Soffer and O. Teryaev, Phys.
Rev. $\bf{D70}$, 116004 (2004).

\bibitem{7} D. Drechsel and L. Tiator Ann. Rev. Nucl. Part. Sci. $\bf{54}$, 69 (2004), nucl-th/0406059;
M. Gorchtein, D. Drechsel, M.M. Giannini, E. Santopinto and L.
Tiator, Phys. Rev. $\bf{C70}$, 055202 (2004) 055202, hep-ph/0404053.

\bibitem{8} B. Badelek, J. Kiryluk, and J. Kwiecinski Phys. Rev. $\bf{D61}$, 014009, hep-ph/9907569;
B. Badelek, J. Kwiecinski and B. Ziaja, Eur. Phys. J. $\bf{C26}$, 45
(2002), hep-ph/0206188.

\bibitem{9} D. Burkert and Z.J. Li, Phys. Rev. $\bf{D47}$, 46 (1993)..

\bibitem{10} V. Bernard, Prog. Part. Nucl. Phys. $\bf{60}$, 82 (2006);
V. Bernard et al., Phys. Rev. $\bf{D67}$, 076008 (2003); X. Ji et
al., Phys. Lett. $\bf{B472}$, 1 (2000).

\bibitem{11} G. Altarelli and G. Parisi, Nucl. Phys. $\bf{B126}$, 298 (1977);
V.N.Gribovand L.N. Lipatov, Sov. J. Nucl. Phys. $\bf{15}$, 438
(1972); Yu.L.Dokshitzer, Sov. Phys. JETP. $\bf{46}$, 641 (1977).

\bibitem{12} X.R. Chen, J.H. Ruan, R. Wang, P.M. Zhang and W. Zhu, Int.
J. Mod. Phys. $\bf{E23}$, 1450057 (2014), hep-ph/1306.1872.

\bibitem{13} W. Zhu and J.H. Ruan, Nucleon spin structure I:
a dynamical determination of polarized gluon distribution in the
proton, this serious works.

\bibitem{14} W. Zhu, Nucl. Phys. $\bf{B 551}$, 245 (1999), hep-ph/9809391;
W. Zhu, J.H. Ruan, Nucl. Phys. $\bf{B559}$, 378 (1999),
hep-ph/9907330v2.

\bibitem{15} W. Zhu, Z.Q. Shen and J.H. Ruan, Nucl.Phys. $\bf{B692}$, 417 (2004), hep-ph/0406212v2.

\bibitem{16} W. Zhu and J.H. Ruan, Nucleon spin structure II: spin dependent structure function
$g_1^p$ at small $x$, this serious works.

\bibitem{17} J. Sakurai, Currents and Mesons (The University of
Chicago Press, Chicago, 1969);

\bibitem{18} HERMES Collaboration, A. Airapetian et al., Eur. Phys. J. C26
(2003) 527, hep-ex/0210047; Phys. Rev., $\bf{D75}$, 012007 (2007).

\bibitem{19} E143 Collaboration, K. Abe et al., Phys. Rev. Lett. $\bf{78}$, 815 (1997),
hep-ex/9701004.

\bibitem{20} CLAS Collaboration, R. Fatemi et al., Phys. Rev. Lett. $\bf{91}$, 222002 (2003),
nucl-ex/0306019;  K. V. Dharmawardane et al., Phys. Lett. $\bf{B
641}$, 11 (2006); P. E. Bosted et al., Phys. Rev. $\bf{C75}$, 035203
(2007); Y. Prok et al., Phys. Lett. $\bf{B672}$, 12 (2009).

\bibitem{21} E.D. Bloom and E.J. Gilman, Phys. Rev. Lett. $\bf{25}$, 1140 (1970); Phys.
Rev. $\bf{D4}$, 2901 (1971).

\bibitem{22} M.Bacchi, L.Feretti, G.Giovannini, F.Govoni, Astron. Astrophys.
$\bf{400}$, 465 (2003),  hep-ph/0301206.

\bibitem{23} M. Gl$\ddot{u}$ck, E. Reya, and A. Vogt, Eur. Phys.
J. $\bf{C5}$, 461 (1998); M. Gl$\ddot{u}$ck, E. Reya, M. Stratmann,
W. Vogelsang, Phys.Rev. $\bf{D63}$, 094005 (2001).




\end{thebibliography}
\end{document}